\documentstyle[aps,preprint]{revtex}

\begin{document}

\title{Generalized seniority scheme in light Sn isotopes }

\author {N.\ Sandulescu $^{a,b}$, J.\ Blomqvist $^b$,T.\ Engeland $^c$, 
	M.\ Hjorth-Jensen $^d$, A.\ Holt $^c$ , R.\ J.\ Liotta $^b$ and 
        E.\ Osnes $^c$ }

\address{$^a$Institute of Atomic Physics, P.O.Box MG-6, Bucharest, Romania}

\address{$^b$Royal Institute of Technology, Physics Department
Frescati, S-10405, Stockholm,Sweden}

\address{$^c$Department of Physics, 
University of Oslo, N-0316 Oslo, Norway}

\address{$^d$Nordita, Blegdamsvej 17, DK-2100 K\o benhavn \O, Denmark}

\maketitle

\begin{abstract}
The  yrast generalized seniority  states  are compared with  the   
corresponding  shell model  states for the case of the Sn
isotopes $^{104-112}$Sn. For most of the cases the energies 
agree within 100 keV and the overlaps of the wave
functions are greater than 0.7
\end{abstract}

In the last years the region of light Sn isotopes has been intensively 
investigated both  from experimental and theoretical perspectives.
The main goal has been  to study the excitation mechanisms
around  the exotic isotope $^{100}$Sn,
the heaviest symmetric double magic nucleus recently
produced in nuclear fragmentation reactions \cite{Sn100a,Sn100b}. 

The simplest approach in analysing the spectra of light Sn isotopes is
to consider $^{100}$Sn as an inert core and to treat only  neutron degrees
of freedom, using the single-particle orbits
of the $N=50-82$ shell as model space, i.e.\ the orbits
$0g_{7/2}$, $1d_{5/2}$, $1d_{3/2}$, $2s_{1/2}$ and
$0h_{11/2}$. Extensive shell model
calculations have been performed along this line \cite{oslo95a}. 
Using a 
Lanczos iteration method states for as many as 12  
extra-core neutrons have been calculated. 
Similar studies have also been done in heavy Sn isotopes \cite{oslo95c}
and in the $N=82$ isotones \cite{oslo96d}, where systems with up to 
14 valence particles have been studied.
On the other hand, a large part of the spectra of light Sn
isotopes can be rather well  described in terms of selected 
configurations such as 
those represented by simple 
quasiparticle excitations  \cite{Stockholm95}.
Therefore, one expects that at least  a part of the low-lying states 
in this region can be approximated by shell-model subspaces with 
reduced dimensions.
One  alternative in truncating the shell model space to 
smaller spaces is offered by 
the generalized seniority scheme GSEN \cite{Talmi71}.
In the mass region of Sn isotopes  GSEN was
applied many years ago \cite{Ottaviani69,Gambhir71}, but then for  
heavier isotopes.
Because complete  shell model calculations were difficult to perform by
that time,  the GSEN results were compared with the ones given by the 
quasiparticle 
Tamm-Dancoff  approximation (QTD) \cite{Gambhir71}. 
It was concluded  that  GSEN and QTD gave similar spectra,
with differences which were  in general less than 100 keV  \cite{Gambhir71}.
Later the admixture of seniority four
states into seniority zero and two states
was analysed \cite{Bonsignori85,Allaart88}. It was  found that 
for some states  the  admixture from seniority four states
could be as large as  $20\%$ .

The aim of the present work is to analyse the accuracy of  the GSEN scheme
for the case of light Sn isotopes. Here we take  advantage of  the fact 
that we can perform complete  shell model 
calculations \cite{oslo95a} 
and thus  exactly check the accuracy of the GSEN 
truncation.
 
One could have a first indication about the validity of the GSEN scheme 
by analysing the experimental binding energies (B.E.) as a function of the
number of neutron pairs, $n$. 
In GSEN this dependence is given by  \cite{Talmi71}
\begin{equation}\label{eq:be}
     B.E.(n) = B.E.(^{100}Sn)+nV_0 + {n(n-1)\over 2\delta}.
\end{equation}
If we fix the parameters $V_0$ and $\delta$ from
 $^{106}$Sn and $^{108}$Sn, the binding energies for A=104 and 110
would be  predicted within 90 and 440 keV, respectively. Considering
the large uncertainties  for the extrapolated B.E.\ 
of $^{100}$Sn  \cite{Wapstra93}, one should take these
estimates as orientative only. 
Nevertheless, they  may indicate that
the generalized seniority zero state
\begin{equation}\label{eq:sen0}
 (S^+)^{n/2} \left | 0\right \rangle , ~~~~  
        S^+ = \sum_{j} C_j(a^+_j a^+_j)_{J = 0},
\label{eq2}    
\end{equation}
could provide a reasonable approximation of the exact shell-model 
ground state.
In Eq.\ (2)  $a^+_j$ denotes 
the particle creation operator.
Two versions of GSEN  have been analyzed. In version~I the amplitudes $C_j$,
which give the distribution of the pairs on the various single-particle orbits, 
are fixed such that the seniority zero state in Eq.~(\ref{eq2}) reproduces the two-particle
shell model state for the ground state of $^{102}$Sn . 
These values are then used  throughout the isotopes from
$^{104}$Sn to $^{112}$Sn. Such an  approach with the constant pair structure 
is within the philosophy
of the original generalized seniority scheme GSEN \cite{Talmi71}. As a simple 
extension called version~II  the amplitudes $C_j$, 
are determined  by minimizing 
the expectation value of the Hamiltonian in the state $ (S^+)^{n/2} \left | 0\right \rangle$,
see Eq.~(2),
for each system separately. This allows the pair structure to change
as a function of the number of particles. 

The validity of seniority schemes have usually been analysed
with  Hamiltonians defined through 
effective interactions fitted to the experimental data. 
In such cases conclusions about the validity of the truncation is 
affected by the fact that the interaction is renormalized as to include 
the effects of the truncation, which is just what we want to estimate.
Thus in the present calculations we use  a microscopically derived effective
interaction to describe the Hamiltonian, using the perturbative many-body
techniques described in Ref.\ \cite{oslo95b}.
In brief,
the derivation of the effective interaction is a three-step process.
First, one needs a free NN interaction $V$ which is
appropriate for nuclear physics at low and intermediate energies. At
present, a meson-exchange picture for the potential model seems to offer a
viable approach. Among such meson-exchange
models one of the most successful is the one-boson-exchange model of the
Bonn group \cite{mac89}. As a starting point for our perturbative analysis 
we use the parameters of the Bonn B potential defined in table A.1 of 
Ref.\ \cite{mac89}. However, 
in nuclear many-body calculations the first problem one is
confronted with is the fact that the repulsive core of the NN potential $V$
is unsuitable for perturbative approaches. This problem is overcome by 
the next
step in our many-body scheme, namely 
by introducing the reaction matrix $G$. Here we calculate the 
$G$-matrix  using the so-called 
double-partitioning 
scheme \cite{oslo95b}.
The single-particle wave functions were chosen to be harmonic oscillator 
eigenstates with the oscillator energy 
$\hbar\Omega = 45A^{-1/3} - 25A^{-2/3}=8.5 $ MeV,  for $A=100$.
The last step consists in defining a
two-body interaction in terms of the $G$-matrix including all diagrams
to third-order in perturbation theory and summing so-called 
folded diagrams to infinite order, see Ref.\ \cite{oslo95b}.
The single-particle
energies for the orbits $1d_{5/2}$, $0g_{7/2}$, $1d_{3/2}$, $2s_{1/2}$
and  $0h_{11/2}$ were fixed as to reproduce the experimental low-lying
states of $^{111}$Sn  \cite{Stockholm95}.

Another property of the Sn isotopes used to justify the 
GSEN approximation is the well-known experimental feature of the 
near constant  spacing between the ground state and the 
first excited $2^+$ states. This  
indicates that the $2^+$ states  may be described as one
broken pair upon 
a  ground state condensate of $0^+$ pairs. Actually one expects a whole group of
low-lying excitations to be expressed as generalized seniority two
 states  \cite {Talmi71} 
\begin{equation}\label{eq:sen2}
    \left | J\right \rangle = D_{J}^+ (S^+)^{n-1}\left | 0 \right \rangle,
\end{equation}
where
\begin{equation}\label{eq;d}
  D^+(J)= \sum_{j_1j_2} X(j_1,j_2;J)(a_{j_1}^+a_{j_2}^+)_{J}.
\end{equation}
In order to investigate these features again two versions of the GSEN are calculated. In version~I
the amplitudes $X(j_1,j_2;J)$ in the two-particle operators
$D^+$ are adjusted to reproduce the corresponding two-particle shell model state
in $^{102}$Sn whereas version~II are found by  by diagonalizing the given interaction
in the space of all possible seniority two basis states, again
for each system separately, In this way one allows
the dynamics to build up the intrinsic structure of the $ D_{J}^+$ operators 
as more pairs are added.

It is worthwhile to stress that the validity of a truncation scheme depends
on the effective interaction employed to describe 
the system. For instance, the validity
of Eq.\ (1) depends on how well the given interaction 
satisfies the relation  \cite{Talmi71}
\begin{equation}\label{eq:int}
   [[H,S^+],S^+]=const.(S^+)^2.    
\end{equation}
The results for the excitation energies of the yrast states are shown in 
Table I. One notices a rather good agreement between the shell-model calculation
and the two versions of GSEN for many of the isotopes. Up to the eight particle case 
the agreement is reasonably  good in both versions, especially in view of the simple model used for 
the pair states compared to the very large shell basis.  As an example, in $^{110}$Sn the number 
of SM basis states for the $2^+$ states is 86990, which should be compared with 9 in the 
GSEN calculation. Above eight particles the deviations start to become significant, 
particularly in version I with fixed pair structures. The version II includes some higher
order pair effects by dynamical changes in the $C_j$ and $X(j_1,j_2:J)$ coefficients, 
but still deviations are up to 0.5~MeV in the worst cases. However,in conclusion
such model calculations as GSEN version I and II show reasonable agreement with the 
shell model ``experimental data''. 

The next information of interest is the properties of the wave functions.
These are analyzed through the overlap squared of the generalized seniority states 
with the exact shell model eigenstate defined by 
\begin{equation}
 \left |\left\langle SM(n,J = 0)\right |(S^{+})^{(n)}\left |0 \right\rangle \right |^2 
             \;\;\;  \mbox{and} \;\;\;
 \left | \left\langle SM(n,J)\right |D_{J}^{+}(S^{+})^{(n - 1)}\left |0\right\rangle \right |^2.
   \label{eq6}
\end{equation}
The results are presented in Table~II. 
Compared to the reasonably good agreement between SM and the GSEN version I and II found for the
energies the wave functions show clear deviations.
For $^{104}$Sn the differences are between 5 and  10~\% whereas in 
$^{112}$Sn the differences have increased to $\approx 85$~\% in some cases in GSEN version I. 
Even the first excited $2^{+}$ state deviates by $\approx 60$~\% in spite of the fact that
this state is well separated from neighbouring nonyrast states which could produce mixing.
So a fixed pair structure description is not meaningful for the heavy Sn isotopes. 

Clear improvement is found in version II. This means that the SM wave
functions contain  important admixtures of other types of configurations
than the seniority zero and two components of the GSEN scheme. 
One may expect that the most important
additional contributions come from  seniority four states, as in the case
of heavier Sn isotopes \cite{Bonsignori85,Allaart88}.
As already pointed out in Refs.\ \cite{Bonsignori85,Allaart88} these 
admixtures could be rather important when observables like transition probabilities are 
calculated.

A complete shell model calculation for the whole chain of isotopes from $^{102}$Sn to
$^{130}$S is difficult. The GSEN model has been thought of as a promising approximation. However, 
the present calculation shows that a model space with pairs of seniority zero and two is
 too small.
Configurations with seniority four and probably six will be necessary for a reasonable description. 
Such work is in progress.

This work has been supported by the 
Nordic Academy for Advanced Studies (NorFA) and the Research Council of
Norway (NFR) under the Supercomputing programme. This work was initiated
when one of us, M.\ H.\ J.\ was at the European Centre for Theoretical
Studies in Nuclear Physics and Related Areas, Trento, Italy. Support from 
the Istituto Trentino di Cultura is acknowledged.

\begin{table}[htbp]
\begin{center}
\caption{Yrast low-lying states for $^{104-112}$Sn. Energies
are given in Mev. For the two versions see the text after Eq.~(\ref{eq2}).}
\begin{tabular}{cccccccccccccccc}
 & \multicolumn{3}{c}{$^{104}$Sn} & \multicolumn{3}{c}{$^{106}$Sn} & 
\multicolumn{3}{c}{$^{108}$Sn} & \multicolumn{3}{c}{$^{110}$Sn} & 
\multicolumn{2}{c}{$^{112}$Sn} \\
 $J^{\pi}$&SM&I&II&SM&I&II&SM&I&II&SM&I&II&SM&I&II \\
\hline
$2^{+}_{1}$ & 1.45 & 1.51 & 1.53 &
                1.42 & 1.56 & 1.54 &
                1.57 & 1.80& 1.64 & 1.63 & 2.17& 1.71 & 1.65 & 1.65& 1.72\\
$4^{+}_{1}$ & 1.98 & 2.03 & 2.05 &
                2.13 & 2.21& 2.25 &
                2.34 &2.57 & 2.42 & 2.43 & 3.06& 2.64& 2.79 & 2.46& 2.77\\
$6^{+}_{1}$ & 2.22 & 2.23 & 2.22 &
              2.36 & 2.47& 2.43 &

                2.45 &2.74 & 2.67  & 2.73 & 3.09& 2.98 & 2.96 & 2.50& 3.29
\end{tabular}
\end{center}
\label{tab-1}
\end{table}

\begin{table}[htbp]
\begin{center}
\caption{ The overlaps square  of the generalized seniority wave functions
with the corresponding shell model states for different angular momenta.
For the two versions see the text after Eq.~(\ref{eq2}).}
\begin{tabular}{cccccccccccc}
& {$J^{\pi}_i$} & \multicolumn{2}{c}{$^{104}$Sn} & 
\multicolumn{2}{c}{$^{106}$Sn} & \multicolumn{2}{c}{$^{108}$Sn} & 
\multicolumn{2}{c}{$^{110}$Sn} & \multicolumn{2}{c}{$^{112}$Sn} \\
& & I & II & I & II & I & II & I & II & I & II \\
\hline
& $0^{+}_{1}$ & 0.950 & 0.966 & 0.876 & 0.938 & 0.796 & 0.924 &
                0.742 & 0.905 & 0.767& 0.909 \\
& $2^{+}_{1}$ & 0.931 & 0.927 & 0.787 & 0.815 & 0.663 & 0.780 & 
                0.438 & 0.790 & 0.420 & 0.776 \\
& $4^{+}_{1}$ & 0.906 & 0.906 & 0.798 & 0.821 & 0.482 & 0.743 & 
                0.236 & 0.764 & 0.173& 0.680 \\
& $6^{+}_{1}$ & 0.918 & 0.943 & 0.817& 0.895 & 0.660 & 0.794 & 
                0.401 & 0.739 & 0.167& 0.695 \\
\end{tabular}
\end{center}
\label{tab-2}
\end{table}


\begin{references}
\bibitem{Sn100a}M.\ Lewitowicz {\em et al.}, 
Phys.\ Lett.\ {\bf B332}, 20 (1994).
\bibitem{Sn100b}R.\ Schneider {\em et al.}, 
Zeit.\ Phys., {\bf A348}, 241 (1994).
\bibitem{oslo95a} T.\ Engeland, M.\ Hjorth-Jensen, A.\ Holt and E.\ Osnes, 
 Phys.\ Scripta {\bf T56}, 58 (1995).
\bibitem{oslo95c} T.\ Engeland, M.\ Hjorth-Jensen, A.\ Holt,
and E.\ Osnes, in proceedings of the International Workshop 
on Double-beta Decay, 
(World Scientific, Singapore, 1996), p.\ 421.
\bibitem{oslo96d}A.\ Holt, T.\ Engeland, 
M.\ Hjorth-Jensen, E.\ Osnes and J.\ Suhonen,
submitted to Nucl.\ Phys.\ {\bf A}.
\bibitem{Stockholm95} N.\ Sandulescu, A.\ Blomqvist and R.\ J.\ Liotta,
Nucl.\ Phys.\ {\bf A582}, 257 (1995).
\bibitem{Talmi71}I.\ Talmi, Nucl.\ Phys.\ {\bf A172}, 1 (1971).
\bibitem{Ottaviani69}P.\ L.\ Ottaviani and M.\ Savoia, Phys.\ Rev.\
{\bf 187}, 1306 (1969).
\bibitem{Gambhir71}Y.\ K.\ Gambhir, A.\ Rimini and T.\ Weber, 
Phys.\ Rev.\ {\bf C3}, 1965 (1971).
\bibitem{Bonsignori85}G.\ Bonsignori, M.\ Savoia, K.\ Allaart, 
A.\ van Egmont and
G.\ Te Velde, Nucl.\ Phys.\ {\bf A432}, 389 (1985).
\bibitem{Allaart88}K.\ Allaart, E.\ Boeker, G.\ Bonsignori,
M.\ Savoia and Y.\ K.\ Gambhir, 
Phys.\ Rep.\ {\bf 169}, 211 (1988).
\bibitem{Wapstra93}G.\ Audi and A.\ H.\ Wapstra, Nucl.\ Phys.\
{\bf A565}, 1 (1993).
\bibitem{oslo95b} M.\ Hjorth-Jensen, T.\ T.\ S.\ Kuo and 
E.\ Osnes, Phys.\ Rep.\ {\bf 261}, 125 (1995).
\bibitem{mac89} R.\ Machleidt, Adv.\ Nucl.\ Phys.\ {\bf 19}, 189 (1989).

\end{references}
\end{document}